\documentclass{ws-mpla}
\usepackage{color}
\usepackage{graphicx}% Include figure files
\usepackage{dcolumn}% Align table columns on decimal point
\usepackage{bm}% bold math

\newcommand{\dv}{\partial\hspace{-7pt}\slash}

\newcommand{\<}{\langle}
\renewcommand{\>}{\rangle}
\newcommand{\be}{\begin{equation}}
\newcommand{\ee}{\end{equation}}
\newcommand{\bea}{\begin{eqnarray}}
\newcommand{\eea}{\end{eqnarray}}

\begin{document}

\markboth{Xiao-Gang He} {Leptogenesis and LHC Physics with Type III See-Saw}

%%%%%%%%%%%%%%%%%%%%% Publisher's Area please ignore %%%%%%%%%%%%%%
%\catchline{}{}{}{}{}
%%%%%%%%%%%%%%%%%%%%%%%%%%%%%%%%%%%%%%%%%%%%%%%%%%%%%%%%%%%%%%%%%%%

\begin{center}
{\Large {\bf Leptogenesis and LHC Physics \\with Type III See-Saw }\footnote{Presented by Xiao-Gang He. e-mail: hexg@phys.ntu.edu.tw}}
\end{center}

\begin{center}
{Shao-Long Chen$^1$ and Xiao-Gang He$^2$}\\
\vspace*{0.3cm}
$^1$Department of Physics, Maryland University\\College Park, Maryland 20742, USA\\
$^2$Department of Physics,  Center for Theoretical Sciences and LeCosPA Center\\
National Taiwan University, Taipei, Taiwan
\end{center}
%\maketitle

%\pub{Received (Day Month Year)}{}%Revised (Day Month Year)}

%\begin{abstract}
\begin{center}
\begin{minipage}{12cm}

\noindent Abstract:\\
The See-Saw mechanism provides a nice way to explain why neutrino masses are so much lighter than their charged lepton partners. It also provides a
nice way to explain baryon asymmetry in our universe via the leptogenesis mechanism.
In this talk we review leptogenesis and LHC physics in a See-Saw model proposed in 1989, now termed the Type III See-Saw model.
In this model,  $SU(2)_L$ triplet leptons are introduced with the neutral particles of the triplets  playing the role of See-Saw. The triplet leptons have charged partners with standard model gauge interactions resulting in many new features.
The gauge interactions of these particles make it easier for leptognesis with low masses, as low as a TeV is possible. The gauge interactions also make the production and detection of triplet leptons at LHC possible. The See-Saw mechanism and leptogenesis due to Type III See-Saw may be tested at LHC.

%\end{abstract}
\end{minipage}

\end{center}
%\ccode{PACS Nos.: include PACS Nos.}

\setcounter{footnote}{0}
\vskip2truecm

\section{See-Saw Mechanism}

The canonical See-Saw mechanism\cite{seesawI} refers to the neutrino mass matrix of the form
\begin{eqnarray}
{\cal L}_m = - {1\over 2}\left ( \overline \nu^c_L, \overline \nu_R \right ) \left ( \begin{array}{ll} 0& m_D\\ m^T_D& M_R\end{array}\right ) \left ( \begin{array}{l} \nu_L\\ \nu_R^c\end{array}
\right )\;.\nonumber
\end{eqnarray}
If the eigenvalues of $M_R $ are much larger than those of  $m_D$, the light and heavy neutrinos mass matrix are given by
\begin{eqnarray}
m_\nu \approx - m_D M^{-1}_R m_D^T,\;\;\;\;m_N \approx M_R\;.
\end{eqnarray}
For one generation, the light neutrino mass is $m_\nu \approx - (m_D/M_R) m_D$. The suppression factor $m_D/M_R$ makes the light neutrino mass much smaller than
the usual Dirac mass $m_D$ which is of order the charged lepton mass.
This is a very nice way to explain why light neutrino masses are so much lighter than their charged partners.

The lightness of neutrino masses in the See-Saw mechanism is related to a large scale for $M_R$. One
may  ask the question,
how large a scale for $M_R$ needs to be to produce the light neutrino mass of order not larger than an eV. Let us taking some examples of $m_D$ to get a feeling about this.
\begin{eqnarray}
&&m_D = m_e:\;\;\;\;m_{\nu_1} = m_e^2/M_R\;,\;\;\mbox{with}\;\; m_{\nu_1} =0.1\mbox{eV}, M_R \simeq 2.5 \mbox{TeV}\nonumber\\
&&m_D = m_\mu:\;\;\;\;m_{\nu_2} = m_\mu^2/M_R\;,\;\;\mbox{with}\;\; m_{\nu_2} =0.1\mbox{eV}, M_R \simeq 10^8 \mbox{GeV}\nonumber\\
&&m_D = m_\tau:\;\;\;\;m_{\nu_3} = m_\tau^2/M_R\;,\;\;\mbox{with}\;\; m_{\nu_3} =0.1\mbox{eV}, M_R \simeq 3\times 10^{10} \mbox{GeV}.
\end{eqnarray}
If $m_D$ is smaller than $m_e$, $M_R$ can be in the O(100GeV) range.
If $m_D$ is larger than $m_\tau$ or $m_{\nu_i}$ smaller than 0.1 eV,
$M_R$ can be of the grand unification scale $10^{15}$ GeV.

Models with $M_R$ completely fixed in terms of low energy neutrino masses, mixing angle and Majorana phases
can be constructed such that\cite{heavyss}
\begin{eqnarray}
M_R = \hat m_f  U^*_{PNMS} \hat m_\nu^{-1} U_{PMNS} \hat m_f\;,\;\;\;\;f = u,\;\;d,\;\;e.\nonumber
\end{eqnarray}
This type of relation results in the right-handed neutrino masses as low as a few TeV and as high as grand unification scale.

With low See-Saw scale of order a TeV, it gives the hope of testing See-Saw mechanism at the Large Hadron Collider (LHC).
For See-Saw scale in the range from a few TeV to grand unification scale, it is possible to explain why our universe is dominated by matter, the baryon asymmetry problem, through the leptogenesis mechanism\cite{leptogenesis}. A lot of things can be said about the See-Saw mechanism! Whether one can have low mass See-Saw scale which can be probed at LHC and at the same time solve the baryon asymmetry problem depends on the detailed models.

There are different ways to realize See-Saw mechanism. They can be categorized as Type I, Type II and Type III See-Saw mechanisms. The main features of these models are:

{\bf Type I}\cite{seesawI}: Introducing singlet right handed neutrinos $\nu_R:
(1,1,0)$ under the standard model (SM) gauge group, $SU(3)_C\times SU(2)_I \times U(1)_Y$. $\nu_R$ does not have SM gauge interactions. The terms relevant to neutrino masses are given by,
\begin{eqnarray}
{\cal L} = \bar L_L Y_D \tilde H \nu_R + {1\over 2} \bar \nu_R^c M_R \nu_R + h.c.\;,
\end{eqnarray}
where $H\equiv (\phi^+, \phi^0)^T\equiv (\phi^+, (v+h+i
\eta)/\sqrt{2})^T$ is the Higgs doublet, and $\tilde H = i \tau_{2} H^*$.
After the Higgs develops VEV, one obtains the neutrino See-Saw mass matrix,
\begin{eqnarray}
M = \left ( \begin{array}{ll} 0& m_D\\ m^T_D& M_R\end{array}\right ),
\end{eqnarray}
where $m_D = Y_D \<\tilde H\> = Y_D v$.

{\bf Type II}\cite{seesawII}: Introducing a triplet Higgs representation $\Delta:
(1,3,1)$ which leads to neutrino masses through the following terms,
\begin{eqnarray}
{\cal L} = {1\over \sqrt{2}}\bar L^c_L Y_t \Delta L_L\;,\;\;\to m_\nu = Y_t v_t\;,\;\;v_t=\sqrt{2}\<\Delta\> \to {v^2\over M_\Delta}\;\;,\;\;M_\Delta\;\mbox{: mass of}\;\;\Delta.\nonumber
\end{eqnarray}

{\bf Type III}\cite{seesawIII}: Introducing triplet lepton representations $\Sigma: (1,3,0)$ to generate See-Saw neutrino mass matrix. This model was first proposed in 1989.

In this type of model, it is possible to have low See-Saw scale of order a TeV to realize leptogenesis and may have detectable effects at LHC due to
the fact that the heavy triplet leptons have gauge interactions. In the following we describe the model in more detail.

The type III See-Saw model consists, in addition to the SM particles, three left-handed triplets of leptons with zero hypercharge,
\begin{eqnarray}
\Sigma&=&
\left(
\begin{array}{ cc}
   N^0/\sqrt{2}  &   E^+ \\
     E^- &  -N^0/\sqrt{2}
\end{array}
\right), \quad
\Sigma^c=
\left(
\begin{array}{ cc}
   N^{0c}/\sqrt{2}  &   E^{-c} \\
     E^{+c} &  -N^{0c}/\sqrt{2}
\end{array}
\right)\;.
\end{eqnarray}
The renormalizable Lagrangian involving $\Sigma$ is given by
\begin{equation}
\label{Lfermtriptwobytwo}
{\cal L}=Tr [ \overline{\Sigma} i \slash \hspace{-2.5mm} D  \Sigma ]
-\frac{1}{2} Tr [\overline{\Sigma}  M_\Sigma \Sigma^c
                +\overline{\Sigma^c} M_\Sigma^* \Sigma]
- \tilde{H}^\dagger \overline{\Sigma} \sqrt{2}Y_\Sigma L_L
-  \overline{L_L}\sqrt{2} {Y_\Sigma}^\dagger  \Sigma \tilde{H}\, .\nonumber
\end{equation}
Defining $E\equiv E_R^{+ c} + E_R^-$, one obtains the Lagrangian
\begin{eqnarray}
\label{Lfull-ft-2}
{\cal L}&=& \overline{E} i \dv E
+ \overline{N_R^0} i \dv  N^0_R
-  \overline{E}M_\Sigma E -
        \left( \overline{N^{0}_R} \frac{{M_\Sigma}}{2}  N_R^{0c} \,+  \,\text{h.c.}\right)
\nonumber \\
&+&g \left(W_\mu^+ \overline{N_R^0} \gamma_\mu  P_R E
 +  W_\mu^+ \overline{N_R^{0c}} \gamma_\mu  P_L E   \,+  \,\text{h.c.}
 \right) - g\, W_\mu^3 \overline{E} \gamma_\mu  E
 \nonumber\\
\nonumber \\
&-&  \left( \phi^0 \overline{N_R^0} Y_\Sigma \nu_{L}+ \sqrt{2}\phi^0 \overline{E} Y_\Sigma l_{L}
+     \phi^+ \overline{N_R ^0} Y_\Sigma l_{L} - \sqrt{2}\phi^+
 \overline{{\nu_{L}}^c}
Y^{T}_\Sigma E   \,+  \,\text{h.c.}\right)\,.\nonumber
\end{eqnarray}

One can easily identify the terms related to neutrino masses from the above and obtain the mass matrix as
\bea
{\cal L}= -(\overline{\nu_L^c}\,\, \overline{N^{0}} )\left(
\begin{array}{ cc}
  0  &   {Y_\Sigma}^T v/2\sqrt{2} \\
   {Y_\Sigma} v/2\sqrt{2} &  {M_\Sigma}/2
\end{array}
\right)
\left(
\begin{array}{ c}
   \nu_L \\
   N^{0c}
\end{array}
\right)\, .\nonumber
\label{neutralfullmassmatrix}
\eea
This is the standard See-Saw mass matrix.

The charged partners in the triplet $\Sigma$ mix with the SM charged leptons resulting in a mass matrix of the following form
\begin{equation}
{\cal L} = -(\overline{l_R}\,\, \overline{E_R} )
\,\,
\left(
\begin{array}{ cc}
   m_l  &   0 \\
      {Y_\Sigma} v &  {M_\Sigma}
\end{array}
\right) \,\,
\left(
\begin{array}{ c}
   l_L \\
  E_L
\end{array}
\right)\;.
\end{equation}

Diagonalizing the above mass matrices, one can work out the interactions of W, Z, $\gamma$ and Higgs with heavy and light
neutrinos, and also the heavy and light charged leptons. Comparing with Type I See-Saw, the heavy triplet leptons have gauge interactions.
This makes the detection of these particles easier if kinematically allowed at colliders, such as at the LHC.
The gauge interaction of the heavy particles also participate in leptogenesis modifying some properties, comparing with Type I See-Saw.
We now discuss some of the main results related to leptogenesis and LHC physics.

\section{Leptogenesis with Type III See-Saw}

It is a well known fact that our universe is dominated by matter. If the universe started with a symmetric one with equal matter and anti-matter,
why our universe is dominated by matter is the famous  baryon asymmetry problem. Putting things into cosmological perspective, if the universe started with
a symmetric one, at present one would expect that the ratio $n_B/n_\gamma$ of baryon and photon number densities should be
$\sim 10^{-20}$, and also $n_B - n_{\bar B} = 0$.
But observations from Big-Bang nucleosynthesis (BBN) and cosmic microwave background (CMB) imply $
{(n_B - n_{\bar B}) / n_\gamma} = {n_B/ n_\gamma} = (6.15\pm 0.25) \times 10^{-10}$.
Baryon number density is much larger than expectation, ten orders of magnitude larger than expectation!

The leptogenesis\cite{leptogenesis} mechanism is the mechanism that transferring lepton number violation into baryon number violation. The key of the mechanism is that at high enough temperature the baryon number B and lepton number L violating anomalous weak gauge boson interaction with fermions, the sphaleron effect, although not
enough to generated the observed baryon asymmetry, it can change net lepton number into baryon number and then result in baryon asymmetry.
The sphaleraon effect violates $B$ and $L$, but conserves $B - L$.
When sphaleron effect is in effective, $B + L$ is washed out, but $B-L$ is preserved.
If the washout for $B + L$ is complete, for initially non-zero $L_i$, but zero $B_i$,
after the sphaleron effect taken place, the final baryon and lepton numbers $B_f$ and $L_f$ will be given by
\begin{eqnarray}
B_i - L_i = B_f - L_f\;,\;\;\;\;B_f + L_f =0\;, \to B_f = -{1\over 2} L_i\;,\;\; L_f = {1\over 2} L_i\;.\nonumber
\end{eqnarray}
In general the washout is not complete, in the SM with one Higgs doublet: $B_f = -{28\over 79} L_i$.

In order to have leptogenesis to occur, a non-zero lepton number asymmetry must be generated and survive the evolution of the universe to convert to baryon number asymmetry. For this to happen there are 3 Sakharov conditions\cite{sakharov} must be satisfied:
1) Lepton number violation; 2) C and CP violation; and 3) Out of thermal equilibrium when 1) and 2) are in progress.

In the See-Saw models, lepton number is violated by the Majorana mass term for heavy neutrino $N$.
When a heavy neutrino $N$ decays, it generates lepton number asymmetry $\epsilon_N$ if there is CP violation in the interaction.
A net lepton number can then be generated if the heavy neutrino decays occurred out of thermal equilibrium.
With the help of the sphaleron effect, one would have at present\cite{leptogenesisIII}
\begin{eqnarray}
{n_B\over n_\gamma} = -{28\over 79}Y^{eq}_N\epsilon_N \eta \;,
\end{eqnarray}
where $Y^{eq}_N = n^{eq}/s$ with $n^{eq}_N$ be the density of $N$ at equilibrium,
and $s$ be the entropy density.
$\eta$ is the efficiency factor of surviving washout and producing net lepton number
during the evolution. It is determined by solving Boltzmann equation taking into
account processes of lepton number conversing (washout asymmetry)
and lepton number conserving processes which is model dependent.

In Type III See-Saw model, lepton number violation is provided by
$-\frac{1}{2} Tr [\overline{\Sigma}  M_\Sigma \Sigma^c
                +\overline{\Sigma^c} M_\Sigma^* \Sigma]$ term in the lagrangian which violated lepton number by two units.
$N^{0}$ and $E^{\pm}_i$ decays produce lepton number asymmetry.

Lepton number asymmetry $\varepsilon_1$ by lightest heavy neutrino and its charged
partners is given by\cite{leptogenesisIII}
\begin{eqnarray}
\varepsilon_1=-\sum_{j=2,3}\frac{3}{2}
  \frac{M_1 }{M_j }\frac{\Gamma_j }{M_j }
  I_j\frac{2 S_j - V_j}{3} \,,
\end{eqnarray}
  where
\begin{eqnarray}
&&I_j = \frac{ \hbox{Im}\,[ (\lambda  \lambda^\dagger)_{1j}^2 ]}
{|\lambda \lambda^\dagger |_{11} |\lambda \lambda^\dagger |_{jj}}
 \, ,\;\qquad
\frac{\Gamma_j}{M_j} = \frac{|\lambda \lambda^\dagger |_{jj}}{8\pi}
\equiv \frac{\tilde{m}_j M_j}{8\pi v^2}
\,,\nonumber\\
&&S_j = \frac{M^2_j  \Delta M^2_{1j}}{(\Delta M^2_{1j})^2+M_1 ^2
   \Gamma_j ^2} \,, \;\;
V_j = 2 \frac{M^2_j }{M^2_1}
\bigg[ (1+\frac{M^2_j }{M^2_1})\log(1+\frac{M^2_1}{M^2_j })
- 1 \bigg]\,,\nonumber\\
&&\Delta M^2_{ij}=M^2_j-M^2_i\,,\;\;\;\;\;\;\;\;\;\;\;\lambda = Y_\Sigma/\sqrt{2}\;.\label{tildem}
\end{eqnarray}
In the hierarchical limit $M_{2,3}\gg M_1$ of the heavy leptons,
\begin{eqnarray}
\varepsilon_1=\sum_{j=2,3}
\frac{3}{16 \pi} \frac{M_1 }{M_j }
\frac{ \hbox{Im}\,[ (\lambda  \lambda^\dagger)_{1j}^2 ]}
{|\lambda \lambda^\dagger |_{11} }\;.\nonumber
\end{eqnarray}
If heavy neutrinos are quasi-degenerate, for $\Delta M^2_{12} = \Gamma^2_2$ and $M_3 \gg M_{1,2}$,
\begin{eqnarray}
\varepsilon_1 = {3\over 4\pi} \frac{ \hbox{Im}\,[ (\lambda  \lambda^\dagger)_{12}^2 ]}
{|\lambda \lambda^\dagger |_{11} }(\ln2 -1)\;.\nonumber
\end{eqnarray}
In this case it it easy to obtain a large $\varepsilon_1$, even as
large as of order one.

\begin{figure}[t]
$$
\includegraphics[width=12cm]{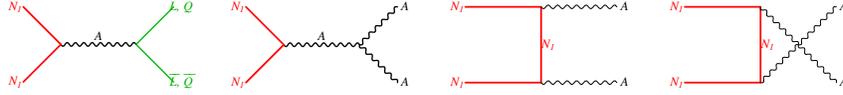}$$
\caption{\label{FeynT}\em
New Feynman diagrams needed to consider when solving Boltzman equation.}
\end{figure}

To obtain the efficiency factor $\eta$, one needs to solve for Boltzmann equation considering relevant processes lepton number violating and conserving ones.
In the type III See-Saw there are new processes to be considered which are shown in Fig. \ref{FeynT} compared with Type I See-Saw.

There are some interesting features in the Type III See-Saw model due to the fact that the heavy leptons have gauge interactions. These gauge interactions can make the heavy lepton
easily be in thermal
equilibrium with other SM particles at high temperature. $Y^{eq}_{N}$ is well determined by counting relativistic particle content in the model. For Type III See-Saw\cite{leptogenesisIII},
\begin{eqnarray}
{n_B\over n_\gamma}= -0.029
\epsilon_1 \eta.
\end{eqnarray}

\begin{figure}[t]
$$
\includegraphics[height=5.4cm]{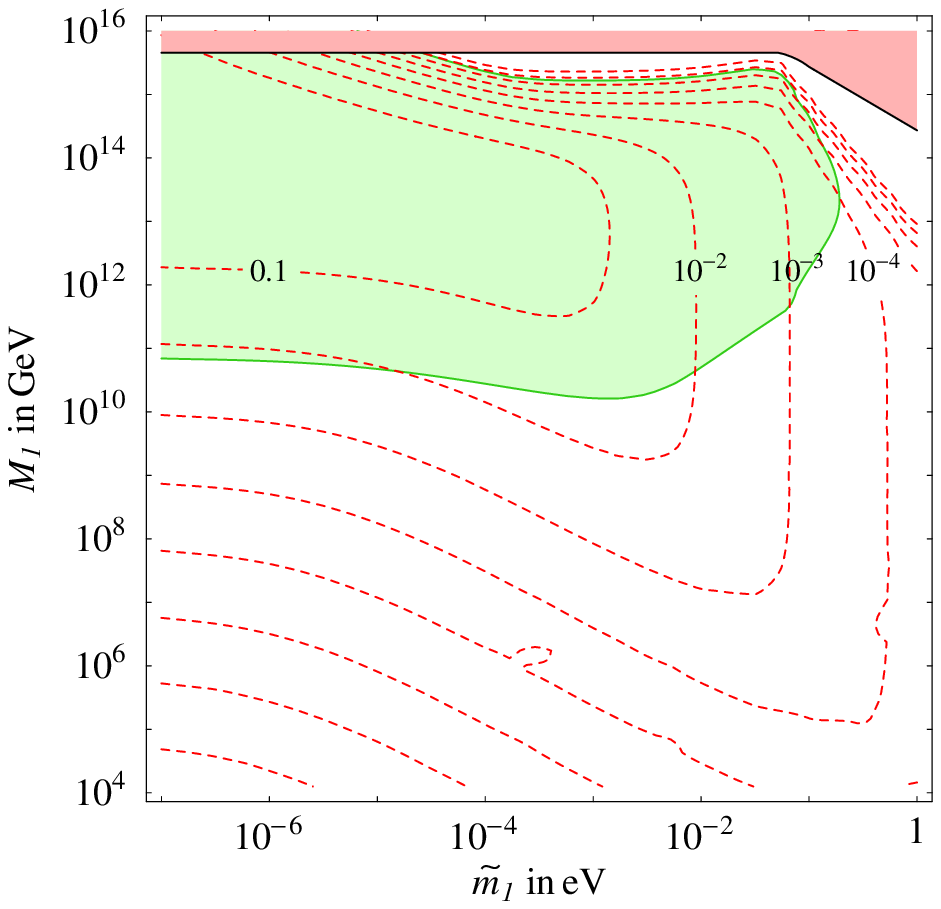}~~~
\includegraphics[height =6cm]{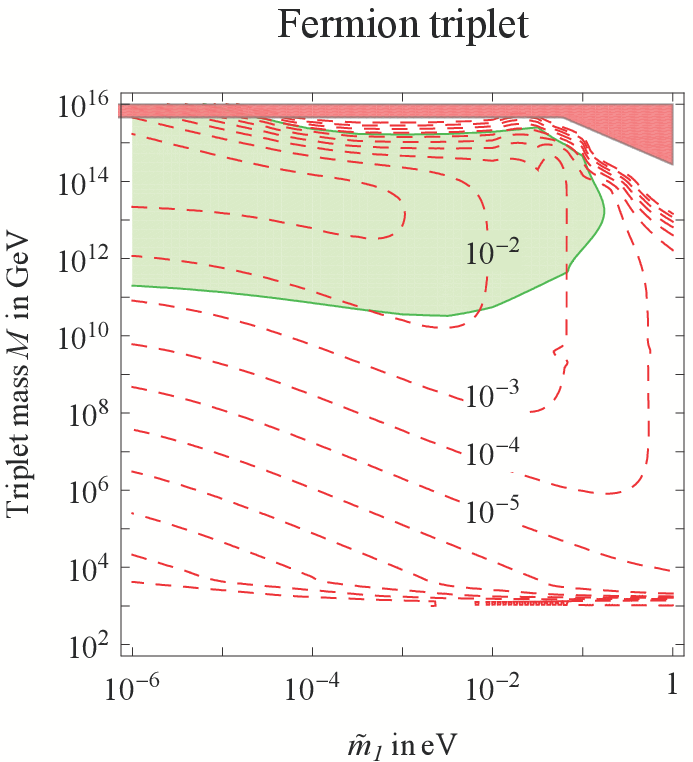}$$
\caption{\label{etaT}\em
Contour-levels of the efficiency $\eta$.
Left: without Sommerfeld effect. Right: with Sommerfeld effect. $\tilde m_1$ is defined by eq.(\ref{tildem}).
}
\end{figure}

At a lower temperature these heavy leptons quickly decouple with others and decay out of thermal equilibrium.
This makes it easier to have large efficiency factor $\eta$ compared
with Type I See-Saw where the heavy neutrinos have no gauge interactions.
In Fig. \ref{etaT} we show the efficiency factor obtained in Refs. \cite{leptogenesisIII} and \cite{leptogenesisIIIa}.

For hierarchical heavy lepton case, in order to reproduce the observed baryon asymmetry, the lightest heavy lepton must satisfy $M_1 > 10^{10}$ GeV.
Leptogenesis works in Type III See-Saw. Although satisfactory for leptogenesis, this scenario makes the See-Saw scale irrelevant for LHC study ad therefore difficult to test the model.
However, in the $M_1$ and $M_2$ degenerate case, CP asymmetry $\epsilon_1$ can be of order one, there is the possibility to allow $M_1$ as low as 1 TeV as can be seen from the left figure in Fig. \ref{etaT}. This makes it possible to test leptogenesis in Type III See-Saw at LHC.

It has recently been point out that charged particle annihilation at non-relativistic speed, due
to the Coulomb interaction, the cross section is enhanced by the Sommerfeld
effect. This effect makes washout stronger leading to the need of a larger
$M_1$ to realize successful leptogenesis. The resulting efficiency factor $\eta$ is shown by the figure on the right in Fig. \ref{etaT}.
It can be seen that $M_1$ is raised to be larger than
a minimal value of\cite{leptogenesisIIIa} $1.6$ TeV. Whether this is a robust lower bound remain to be studied since flavor dependent leptogenesis\cite{flavorlepto} has not been included.
The inclusion of this may lower the $M_1$ mass somewhat.

\section{Type III See-Saw at LHC}

The LHC is now ready to take data. It has the potential to test many of the models beyond the SM.
It is a very exciting period for particle physics. The Type III See-Saw may also have a chance to be
directly tested at LHC by discovering the heavy particles in the triplet leptons. Even more interesting is that
in Type III See-Saw heavy neutrinos as low as a TeV in mass can explain baryon asymmetry in our universe. There is a chance
to directly test leptogenesis hypothesis too.

As far as detection of See-Saw heavy neutrino is concerned, the Type III See-Saw has some advantages compared with Type I See-Saw.
In Type III See-Saw, besides heavy neutral Majorana particle $N^0$, there are
also heavy charged particles $E^\pm$. They have
gauge interactions and are much easier than the singlet heavy neutrino
model to be studied at LHC.

Whether the heavy triplet leptons can be discovered depends on many things. First one must consider the production rate of the heavy triplet leptons.  The main production channels of these heavy particles are
\begin{eqnarray}
q \bar q \to Z^*/\gamma^* \to E^+ E^-\;,\;\;\;\; q \bar q' \to W^*\to E^\pm N^0\;.
\end{eqnarray}
The calculations are straightforward.
The partonic production cross sections, summed over initial state colors and over final state polarizations,
and averaged over initial state polarizations, are\cite{production}
\begin{eqnarray}
\hat\sigma &=&\frac{\beta(3-\beta^2)}{48\pi }\label{eq:sigmatot}
{1\over N_c} \hat s (V_L^2 + V_R^2)
\end{eqnarray}
where $N_c = 3$ is the number of colors, $\beta\equiv \sqrt{1-4M^2/\hat s}$ is the $N$ velocity ($0\le\beta\le 1$) and
\begin{eqnarray}
&&V_a=0  \qquad \hbox{for $q\bar q\to N^0 N^0$ };\nonumber\\
&&V_a= \displaystyle\frac{Q_q e^2}{\hat s}+
\frac{g_A^q g_2^2}{\hat{s} - M_Z^2},
\qquad \hbox{for $q\bar q\to E^+ E^-$ };\nonumber\\
&&V_a = \displaystyle\frac{g_2^2}{\hat s-M_W^2}\frac{\delta_{AL}}{\sqrt{2}}, \qquad\hbox{for $u\bar d\to E^+ N^0$ };
\end{eqnarray}
where $g_a^q = T_3 - s_W^2 Q_q$ is the $Z$ coupling of quark $q$ for $a=\{L,R\}$.

\begin{figure}[t]
$$
\includegraphics[width=7cm]{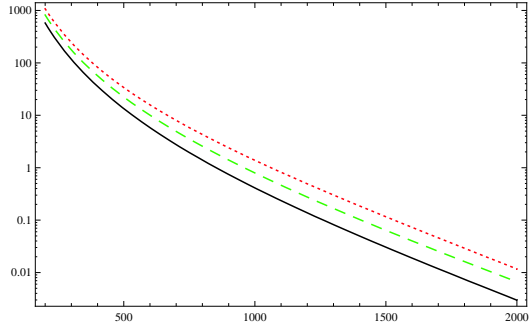}$$
\caption{\label{cross}\em
Production cross sections (in $fb$ unit) as functions of the heavy triplet masses (in GeV unit) for $N^0 E^+$ (doted curve), $E^+E^-$ (dashed curve) and $N^0 E^-$ (solid curve).}
\end{figure}

Since the interaction couplings are the known gauge couplings of the SM, the only unknowns in
the cross sections are the heavy particle masses.
We have calculated the production cross sections for heavy triplet particle masses up to 2 TeV at LHC which are shown in Fig. \ref{cross}.
We see that the cross sections are not negligibly small. It may be able to produce a lot of these particles.

To be sure that the triplet leptons are produced, one has to study the decays to specify detection signals.
The decay modes for the heavy triplet leptons are mainly
\begin{eqnarray}
&&N \to \;\;\;\;l^\pm W^\mp\;,\;\;\;\;\nu Z\;,\;\;\;\;\nu h\;,\nonumber\\
&&E^\pm \to\;\;\;\; l^\pm Z,\;\;\;\;\nu W^\pm\;,\;\;\;\;l^\pm
h\;.\nonumber
\end{eqnarray}

With these decay modes, one can then decide detection signals. This has been discussed in the literature\cite{production,signal}.
Possible signals have been classified according to how many usual charged leptons in the final states. We now list them below\cite{signal}:

i) 6 charged leptons $l^\pm l^\pm l^\pm l^\mp l^\mp l^\mp$:
$E^+ E^-  \to \ell^+ Z \, \ell^- Z$ with $Z \to \ell^+ \ell^-$.

ii) 5 charged leptons $l^\pm l^\pm l^\mp l^\mp l^+$:
$E^+ N \to \ell^+ Z \, \ell^\pm W^\mp$ with $Z \to \ell^+ \ell^-, W \to \ell \nu$;
$E^+ N \to \ell^+ Z\, \nu Z $ with $Z \to \ell^+ \ell^-$.

iii) 4 charged leptons $l^\pm l^\pm l^\pm l^\mp$:
$E^+ N \to \ell^+ Z \, \ell^+ W^-$ with $Z \to \ell^+ \ell^-$, $W \to q \bar q'$,

iv) 4 charged leptons $l^+ l^+ l^- l^-$:
$E^+ E^- \to \ell^+ Z \, \ell^- Z $ with $ZZ \to \ell^+ \ell^- q \bar q / \nu \bar\nu$;
$E^+ E^- \to \ell^+ Z \, \ell^- H / \ell^+ H \, \ell^- Z $ with $Z \to \ell^+ \ell^- , H \to q \bar q$;
$E^+ E^- \to \nu W^+ \ell^- Z / \ell^+ Z \nu W^- $ with $Z \to \ell^+ \ell^- , W \to \ell \nu$;
$ E^\pm N \to \ell^\pm Z \, \ell^- W^+ $  with $ Z \to \ell^+ \ell^- , W \to q \bar q'$.

v) 3 charged leptons $l^\pm l^\pm l^\mp$:
$ E^+ N \to \ell^+ Z \, \ell^\pm W^\mp $
with $Z \to q \bar q / \nu \bar \nu, W \to \ell
  \nu$;
 $ E^+ N \to \ell^+ H \, \ell^\pm W^\mp$ with
  $H \to q \bar q , W \to \ell \nu$;
$E^+ N \to \bar \nu W^+ \, \ell^\pm W^\mp$ with
 $W \to \ell \nu$.

vi) 2 charged leptons $l^\pm l^\pm$: $E^+ N \to \ell^+ Z \, \ell^+ W^-$ with
  $Z \to q \bar q / \nu \bar \nu, W \to q \bar
  q'$;
$E^+ N \to \ell^+ H \, \ell^+ W^- $,
 with $H \to q \bar q , W \to q \bar q'$;
$E^+ N \to \bar \nu W^+ \, \ell^+ W^-$ with
  $W^+ \to \ell \nu, W^- \to q \bar q'$.

vii) 2 charged leptons $l^+l^- jjjj$: $E^+ N \to \ell^+ Z \, \ell^- W^+$ with
  $\to q \bar q / \nu \bar \nu, W \to q \bar
  q'$;
$E^+ N \to \ell^+ H \, \ell^- W^+$ with
  $H \to q \bar q , W \to q \bar q'$;
$E^+ N \to \bar \nu W^+ \, \ell^- W^+$ with
  $W W\to \ell \nu q \bar q'$;
 $E^+ E^- \to \ell^+ Z \, \ell^- Z$ with
  $Z \to q \bar q / \nu \bar \nu$;
 $ E^+ E^- \to \ell^+ Z \, \ell^- H / \ell^+ H \, \ell^- Z$ with
  $Z \to q \bar q / \nu \bar \nu , H \to q \bar
  q$;
$E^+ E^- \to \ell^+ H \, \ell^- H$ with
  $H \to q \bar q$.

viii) 1 charged lepton $l^\pm jjjj$:
$E^+ N \to \bar \nu W^+ \, \ell^\mp W^\pm $ with
  $W \to  q \bar q'$;
$E^+ N \to \ell^+ Z/H \, \nu Z/H $ with $Z \to  q \bar q'/\nu \bar \nu,H \to q \bar
  q$;
$E^+ E^- \to \ell^+ Z/H \, \nu W^-$ with $Z \to q \bar q/\nu \bar \nu,H \to q \bar q,W \to q \bar
  q'$;
$E^+ E^- \to \bar \nu W^+ \, \ell^- Z/H $ with $Z \to q \bar q/\nu \bar \nu,H \to q \bar q,W \to q \bar
  q'$.

Signals from i) and ii) have too small cross sections. They are not good for discovery. Signals from iii) and iv) can provide clean measurement of the heavy triplet masses.
Signals from v) and vi) have excellent final state for the discovery with relatively high signal rate and small background\cite{signal}.
Signals from vii) and viii) have large cross sections, but large background.

Combining all signals, it has been shown\cite{signal} that it is possible to  reach heavy triplet lepton masses up to of order 300 GeV with integrated luminosity of 30 $fb^{-1}$.
Up to 1.5 TeV can be probed with integrated luminosity of 300 $fb^{-1}$. The lower range of leptogenesis in the See-Saw III can be probed.
There is hope that leptogenesis can be tested at LHC. Since to have low mass leptogenesis, large CP violation in heavy neutrino decays are needed, the study of CP violation at LHC should be a good test for leptogenesis.

Of course, one has to be open minded that See-Saw III does not have to do all jobs. If nature choose it to do just the See-Saw mechanism job with a  mass lower than required by leptogenesis, and to be detected at LHC, we are already luck enough.

The $E^-$ in many ways is similar to a 4th generation charged lepton. One can, however, distinguish them by looking at the  branching ratio for $E^\pm \to
l^\pm Z$. If it is a 4th generation charged lepton, the branching ratio is expected to be much smaller than that from the heavy triplet charged lepton.

\section{Conclusions}

There are different ways the See-Saw mechanism can be realized. The Type III See-Saw is a very simple mechanism for light neutrino masses.
Compared with Type I See-Saw, the heavy neutrinos introduced to do the See-Saw job have their charged partners. These particles have SM gauge interactions
resulting in many interesting new features. In this talk we have reviewed some interesting results related to leptogenesis and collider physics at LHC.

The SM gauge interactions enable the heavy triplet particle to be in thermal equilibrium with other SM particle making the initial abundance of the heavy particle much
more certain to calculate. These gauge interactions drop with temperature quickly leading
to out of thermal decay of the lightest heavy triplet particle to easily
produce large enough lepton number violation capable of explaining baryon asymmetry via leptogenesis, and the mass of the heavy neutrino
can be as low as a TeV in the degenerate case. This last possibility also makes it possible to study leptogenesis at LHC.

The production and detection are also much easier than Type I See-Saw due to the existence of gauge interactions. Heavy neutrinos
and their partners as heavy as a TeV can be produced  and detected at LHC when
the integrated luminosity reach 300 fb$^{-1}$. Up to 300 GeV can
be detected with 30fb$^{-1}$ integrated luminosity.

It is interesting to study CP violation at LHC since the low mass leptogenesis requires a large CP violation in heavy neutrino decays. This may provide a unique
test for leptogenesis at LHC.

In this talk, we have not discussed many other interesting features of the Type III See-Saw. One of these is the flavor changing neutral current (FCNC) interactions
in lepton sector. The introduction of triplet leptons will induce FCNC at tree level for charged lepton interactions, but not for quarks. Study of signals of
FCNC in lepton sector and low energy processes may provide useful information to distinguish different models\cite{fcnc,g-2,he-oh}.
There are also several studies regarding
extended model buildings\cite{model} and other related issues\cite{other}. The Type III See-Saw and related models should be studied more theoretically to reveal its interesting features, and of
course experimentally to decide
whether it is relevant to reality.

\noindent{\bf Acknowledgments} This work was supported in part by
NSC and NCTS.

%\tighten


\begin{thebibliography}{0}

\bibitem{seesawI}
P.~Minkowski,
  %``Mu $\to$ E Gamma At A Rate Of One Out Of 1-Billion Muon Decays?,''
  Phys.\ Lett.\ B {\bf 67}, 421 (1977);
  %%CITATION = PHLTA,B67,421;%%
  T.~Yanagida, in {\it Workshop on Unified Theories}, KEK report 79-18 p.95 (1979);
  M.~Gell-Mann, P.~Ramond, R.~Slansky,
  in {\it Supergravity} (North Holland, Amsterdam, 1979)
  eds. P.~van~Nieuwenhuizen, D.~Freedman, p.315;
  S.~L.~Glashow, in {\it 1979 Cargese Summer Institute on Quarks and Leptons} (Plenum Press,
  New York, 1980) eds. M.~Levy, J.-L.~Basdevant, D.~Speiser, J.~Weyers, R.~Gastmans and M.~Jacobs,
  p.687;
  R.~Barbieri, D.~V.~Nanopoulos, G.~Morchio and F.~Strocchi,
 % %``Neutrino Masses In Grand Unified Theories,''
  Phys.\ Lett.\ B {\bf 90}, 91 (1980);
 % %%CITATION = PHLTA,B90,91;%%
  R.~N.~Mohapatra and G.~Senjanovic,
 % %``Neutrino mass and spontaneous parity nonconservation,''
  Phys.\ Rev.\ Lett.\  {\bf 44}, 912 (1980);
 % %%CITATION = PRLTA,44,912;%%
  G.~Lazarides, Q.~Shafi and C.~Wetterich,
 % %``Proton Lifetime And Fermion Masses In An SO(10) Model,''
  Nucl.\ Phys.\  B {\bf 181}, 287 (1981).
 % %%CITATION = NUPHA,B181,287;%%


%\cite{He:2008cd}
\bibitem{heavyss}
  X.~G.~He, S.~S.~C.~Law and R.~R.~Volkas,
  %``Determining the heavy seesaw neutrino mass matrix from low-energy
  %parameters,''
  Phys.\ Rev.\  D {\bf 78}, 113001 (2008)
  [arXiv:0810.1104 [hep-ph]].
  %%CITATION = PHRVA,D78,113001;%%

\bibitem{leptogenesis}
 M.~Fukugita and T.~Yanagida,
  %``Baryogenesis Without Grand Unification,''
  Phys.\ Lett.\ B {\bf 174}, 45 (1986).
  %%CITATION = PHLTA,B174,45;%%

\bibitem{seesawII}
W.~Konetschny and W.~Kummer,
  %``Nonconservation Of Total Lepton Number With Scalar Bosons,''
  Phys.\ Lett.\  B {\bf 70}, 433 (1977);
  %%CITATION = PHLTA,B70,433;%%
%
 T.~P.~Cheng and L.~F.~Li,
  %``Neutrino Masses, Mixings And Oscillations In SU(2) X U(1) Models Of
  %Electroweak Interactions,''
  Phys.\ Rev.\  D {\bf 22}, 2860 (1980);
  %%CITATION = PHRVA,D22,2860;%%
%
 G.~Lazarides, Q.~Shafi and C.~Wetterich,
  %``Proton Lifetime And Fermion Masses In An SO(10) Model,''
 Nucl.\ Phys.\  B {\bf 181}, 287 (1981);
  %%CITATION = NUPHA,B181,287;%%
%
 J.~Schechter and J.~W.~F.~Valle,
  %``Neutrino Masses In SU(2) X U(1) Theories,''
  Phys.\ Rev.\  D {\bf 22}, 2227 (1980);
  %%CITATION = PHRVA,D22,2227;%%
%
 R.~N.~Mohapatra and G.~Senjanovic,
  %``Neutrino Masses And Mixings In Gauge Models With Spontaneous Parity
  %Violation,''
  Phys.\ Rev.\  D {\bf 23}, 165 (1981).
  %%CITATION = PHRVA,D23,165;%%


\bibitem{seesawIII}
R.~Foot, H.~Lew, X.~G.~He and G.~C.~Joshi,
  %``SEESAW NEUTRINO MASSES INDUCED BY A TRIPLET OF LEPTONS,''
  Z.\ Phys.\  C {\bf 44}, 441 (1989).
  %%CITATION = ZEPYA,C44,441;%%

\bibitem{sakharov}
A.D. Sakharov, Pisma Zh. Eksp. Teor. Fiz. {\bf 5}, 32(1967).

\bibitem{leptogenesisIII}

%\cite{Hambye:2003rt}
%\bibitem{Hambye:2003rt}
T.~Hambye, Y.~Lin, A.~Notari, M.~Papucci and A.~Strumia,
  %``Constraints on neutrino masses from leptogenesis models,''
  Nucl.\ Phys.\  B {\bf 695}, 169 (2004)
  [arXiv:hep-ph/0312203].
  %%CITATION = NUPHA,B695,169;%%


\bibitem{leptogenesisIIIa}
%\cite{Strumia:2008cf}
%\bibitem{Strumia:2008cf}
A.~Strumia,
  %``Sommerfeld corrections to type-II and III leptogenesis,''
  Nucl.\ Phys.\  B {\bf 809}, 308 (2009)
  [arXiv:0806.1630 [hep-ph]].
  %%CITATION = NUPHA,B809,308;%%

\bibitem{flavorlepto}
%\cite{Blanchet:2008hg}
%\bibitem{Blanchet:2008hg}
  S.~Blanchet,
  %``A New Era of Leptogenesis,''
  arXiv:0807.1408 [hep-ph];
  %%CITATION = ARXIV:0807.1408;%%
%\cite{JosseMichaux:2008ix}
%\bibitem{JosseMichaux:2008ix}
  F.~X.~F.~Josse-Michaux,
  %``Recent developments in thermal leptogenesis: the role of flavours in
  %various seesaw realisations,''
  arXiv:0809.4960 [hep-ph].
  %%CITATION = ARXIV:0809.4960;%%


\bibitem{production}
%\cite{Franceschini:2008pz}
%\bibitem{Franceschini:2008pz}
R.~Franceschini, T.~Hambye and A.~Strumia,
  %``Type-III see-saw at LHC,''
  Phys.\ Rev.\  D {\bf 78}, 033002 (2008)
  [arXiv:0805.1613 [hep-ph]];
  %%CITATION = PHRVA,D78,033002;%%
  %\cite{Bajc:2007zf}
%\bibitem{Bajc:2007zf}
B.~Bajc, M.~Nemevsek and G.~Senjanovic,
  %``Probing seesaw at LHC,''
  Phys.\ Rev.\  D {\bf 76}, 055011 (2007)
  [arXiv:hep-ph/0703080].
  %%CITATION = PHRVA,D76,055011;%%

\bibitem{signal}
%\cite{delAguila:2008hw}
%\bibitem{delAguila:2008hw}
  F.~del Aguila and J.~A.~Aguilar-Saavedra,
  %``Electroweak scale seesaw and heavy Dirac neutrino signals at LHC,''
  arXiv:0809.2096 [hep-ph].
  %%CITATION = ARXIV:0809.2096;%%

\bibitem{fcnc}
%\cite{Biggio:2008in}
%\bibitem{Biggio:2008in}
 % C.~Biggio,
  %``The contribution of fermionic seesaws to the anomalous magnetic moment of
  %leptons,''
 % Phys.\ Lett.\  B {\bf 668}, 378 (2008)
 % [arXiv:0806.2558 [hep-ph]].
  %%CITATION = PHLTA,B668,378;%%
%\cite{Abada:2007ux}
%\bibitem{Abada:2007ux}
A.~Abada, C.~Biggio, F.~Bonnet, M.~B.~Gavela and T.~Hambye,
  %``Low energy effects of neutrino masses,''
  JHEP {\bf 0712}, 061 (2007)
  [arXiv:0707.4058 [hep-ph]];
  %%CITATION = JHEPA,0712,061;%%
  %\cite{FernandezMartinez:2007ms}
%\bibitem{FernandezMartinez:2007ms}
  E.~Fernandez-Martinez, M.~B.~Gavela, J.~Lopez-Pavon and O.~Yasuda,
  %``CP-violation from non-unitary leptonic mixing,''
  Phys.\ Lett.\  B {\bf 649}, 427 (2007)
  [arXiv:hep-ph/0703098].
  %%CITATION = PHLTA,B649,427;%%



\bibitem{g-2}
%\cite{Biggio:2008in}
%\bibitem{Biggio:2008in}
C.~Biggio,
  %``The contribution of fermionic seesaws to the anomalous magnetic moment of
  %leptons,''
  Phys.\ Lett.\  B {\bf 668}, 378 (2008)
  [arXiv:0806.2558 [hep-ph]];
  %%CITATION = PHLTA,B668,378;%%
%\cite{Chao:2008iw}
%\bibitem{Chao:2008iw}
  W.~Chao,
  %``The Muon Magnetic Moment in the TeV Scale Seesaw Models,''
  arXiv:0806.0889 [hep-ph].
  %%CITATION = ARXIV:0806.0889;%%

\bibitem{he-oh} Xiao-Gang He and Sechule Oh, in preparation.


\bibitem{model}
%\cite{Hirsch:2008mg}
%\bibitem{Hirsch:2008mg}
  M.~Hirsch, S.~Morisi and J.~W.~F.~Valle,
  %``Modelling tri-bimaximal neutrino mixing,''
  arXiv:0810.0121 [hep-ph];
  %%CITATION = ARXIV:0810.0121;%%
%\cite{Ma:2008uz}
%\bibitem{Ma:2008uz}
E.~Ma,
  %``Neutrino Mass Seesaw Version 3: Recent Developments,''
  arXiv:0810.5574 [hep-ph];
  %%CITATION = ARXIV:0810.5574;%%
  %\cite{Perez:2007rm}
%\bibitem{Perez:2007rm}
  P.~Fileviez Perez,
  %``Renormalizable Adjoint SU(5),''
  Phys.\ Lett.\  B {\bf 654}, 189 (2007)
  [arXiv:hep-ph/0702287];
  %\cite{Adhikari:2008uc}
%\bibitem{Adhikari:2008uc}
  R.~Adhikari, J.~Erler and E.~Ma,
  %``Seesaw Neutrino Mass and New U(1) Gauge Symmetry,''
  arXiv:0810.5547 [hep-ph].
  %%CITATION = ARXIV:0810.5547;%%
  %%CITATION = PHLTA,B654,189;%%
  %\cite{Dorsner:2006fx}
%\bibitem{Dorsner:2006fx}
  I.~Dorsner and P.~Fileviez Perez,
  %``Upper Bound on the Mass of the Type III Seesaw Triplet in an SU(5) Model,''
  JHEP {\bf 0706}, 029 (2007)
  [arXiv:hep-ph/0612216];
  %%CITATION = JHEPA,0706,029;%%
%\cite{Bajc:2006ia}
%\bibitem{Bajc:2006ia}
  B.~Bajc and G.~Senjanovic,
  %``Seesaw at LHC,''
  JHEP {\bf 0708}, 014 (2007)
  [arXiv:hep-ph/0612029];
  %%CITATION = JHEPA,0708,014;%%
%\cite{Ma:2002pf}
%\bibitem{Ma:2002pf}
  E.~Ma and D.~P.~Roy,
  %``Heavy triplet leptons and new gauge boson,''
  Nucl.\ Phys.\  B {\bf 644}, 290 (2002)
  [arXiv:hep-ph/0206150];
  %%CITATION = NUPHA,B644,290;%%
%\cite{Ma:2002nn}
%\bibitem{Ma:2002nn}
  E.~Ma,
  %``Neutrino mass from triplet and doublet scalars at the TeV scale,''
  Phys.\ Rev.\  D {\bf 66}, 037301 (2002)
  [arXiv:hep-ph/0204013];
  %%CITATION = PHRVA,D66,037301;%%
%\cite{Ma:2001kg}
%\bibitem{Ma:2001kg}
  E.~Ma,
  %``New gauge symmetry of quarks and leptons,''
  Mod.\ Phys.\ Lett.\  A {\bf 17}, 535 (2002)
  [arXiv:hep-ph/0112232].
  %%CITATION = MPLAE,A17,535;%%

\bibitem{other}
%\cite{Chakrabortty:2008zh}
%\bibitem{Chakrabortty:2008zh}
  J.~Chakrabortty, A.~Dighe, S.~Goswami and S.~Ray,
  %``Renormalization group evolution of neutrino masses and mixing in the
  %Type-III seesaw mechanism,''
  arXiv:0812.2776 [hep-ph];
  %%CITATION = ARXIV:0812.2776;%%
%\cite{Mohapatra:2008wx}
%\bibitem{Mohapatra:2008wx}
  R.~N.~Mohapatra, N.~Okada and H.~B.~Yu,
  %``$\nu$-GMSB with Type III Seesaw and Phenomenology,''
  Phys.\ Rev.\  D {\bf 78}, 075011 (2008)
  [arXiv:0807.4524 [hep-ph]].
  %%CITATION = PHRVA,D78,075011;%%





\end{thebibliography}
\end{document}